\documentclass[twocolumn, twocolappendix]{aastex6}

\newcommand\aastex{AAS\TeX}

\newcommand\abs[1]{\ensuremath{\lvert#1\rvert}} 

\shorttitle{\aastex VIP package for high-contrast direct imaging}
\shortauthors{Gomez Gonzalez et al.}

\usepackage{natbib}
\usepackage{graphicx} 	
\usepackage{amsmath}	
\usepackage{amssymb}	
\usepackage{bm}			
\usepackage{txfonts}

\usepackage[utf8]{inputenc}

\usepackage[english]{babel}
\usepackage{mathtools} 
\DeclareMathOperator*{\argmin}{argmin}
\DeclareMathOperator*{\rank}{rank}
\usepackage{gensymb}

\usepackage{listings}
\definecolor{lbcolor}{rgb}{0.9,0.9,0.9}
\usepackage{fnpct} 

\newcommand{\vip}{\texttt{VIP}}
\newcommand{\python}{\texttt{python}}
\newcommand{\git}{\texttt{git}}

\usepackage{enumitem}
\setlist{nosep} 

\begin{document}

\title{\texttt{VIP:} Vortex Image Processing package for high-contrast direct imaging}

\author{Carlos Alberto Gomez Gonzalez, Olivier Wertz, Olivier
	    Absil\altaffilmark{1}, Valentin Christiaens\altaffilmark{2,3} and 
	    Denis Defrère}
\affil{Space sciences, Technologies \& Astrophysics Research (STAR) Institute, Université de Liège, Allée du Six Août 19c, B-4000 Liège, Belgium}
\author{Dimitri Mawet\altaffilmark{4}} 
\affil{Department of Astronomy, California Institute of Technology, Pasadena, CA 91125, USA}
\author{Julien Milli} 
\affil{European Southern Observatory, Santiago, Chile }
\author{Pierre-Antoine Absil} 
\affil{Department of Mathematical Engineering, Université catholique de Louvain, B-1348 Louvain-la-Neuve, Belgium}
\author{Marc Van Droogenbroeck} 
\affil{Institut Montefiore, Université de Liège, B-4000 Liège, Belgium}
\author{Faustine Cantalloube\altaffilmark{5}} 
\affil{CNRS, IPAG, F-38000 Grenoble, France}
\author{Philip M. Hinz} 
\affil{Steward Observatory, Department of Astronomy, University of Arizona, 933 N. Cherry Ave, Tucson, AZ 85721, USA}
\author{Andrew J. Skemer} 
\affil{University of California, Santa Cruz, 1156 High St. Santa Cruz, CA 95064, USA}
\author{Mikael Karlsson} 
\affil{Department of Engineering Sciences, {\AA}ngstr\"{o}m Laboratory, Uppsala University, Box 534, SE-751 21 Uppsala, Sweden }
\author{Jean Surdej\altaffilmark{6}} 
\and
\affil{Space sciences, Technologies \& Astrophysics Research (STAR) Institute, Université de Liège, Allée du Six Août 19c, B-4000 Liège, Belgium} 

\altaffiltext{1}{F.R.S.-FNRS Research Associate}
\altaffiltext{2}{Departamento de Astronomía, Universidad de Chile, Santiago, Chile}
\altaffiltext{3}{Millennium Nucleus “Protoplanetary Disks”, Chile}
\altaffiltext{4}{Jet Propulsion Laboratory, 4800 Oak Grove Drive, Pasadena CA 91109, USA}
\altaffiltext{5}{Office National d’Etudes et de Recherches Aérospatiales (ONERA), Optics Department, BP 72, 92322 Ch\^{a}tillon, France}
\altaffiltext{6}{Honorary F.R.S.-FNRS Research Director}

\begin{abstract}   
	We present the Vortex Image Processing (\vip) library, a \python\ package
	dedicated to astronomical high-contrast imaging. Our package relies
	on the extensive \python\ stack of scientific libraries and aims to provide
	a flexible framework for high-contrast data and image processing.  
	In this paper, we describe the capabilities of \vip\ related to processing
	image sequences acquired using the angular differential imaging (ADI) 
	observing technique. \vip\ implements functionalities for building
	high-contrast data processing pipelines, encompassing pre- and
	post-processing algorithms, potential sources position and flux estimation, 
	and sensitivity curves generation. Among the reference point-spread
	function subtraction techniques for ADI post-processing, \vip\ includes
	several flavors of principal component analysis (PCA) based algorithms,
	such as annular PCA and incremental PCA algorithm capable of processing big
	datacubes (of several gigabytes) on a computer with limited memory. Also,
	we present a novel ADI algorithm based on non-negative matrix factorization
	(NMF), which comes from the same family of low-rank matrix approximations 
	as PCA and provides fairly similar results. We showcase the ADI 
	capabilities of the \vip\ library using a deep sequence on HR8799 taken
	with the LBTI/LMIRCam and its recently commissioned L-band vortex
    coronagraph. Using \vip\ we investigated the presence of additional
    companions around HR8799 and did not find any significant additional point
    source beyond the four known planets. 
    \vip\ is available at \url{http://github.com/vortex-exoplanet/VIP} and is
    accompanied with Jupyter notebook tutorials illustrating the main
    functionalities of the library. 
\end{abstract}

\keywords{Methods: data analysis - Techniques: high angular resolution - 
Techniques: image processing - Planetary systems - Planets and satellites: detection}
                        
\section{Introduction}  
	
	The field of exoplanets is presently one of the most active areas of modern
	astrophysics. In only two decades of exoplanetary science, we count more
	than three thousand confirmed discoveries,
	most of which have been made possible thanks to indirect methods 
	\citep{pepe14}. Only a
	few tens of exoplanets could be directly resolved through high-contrast 
	imaging. The task of finding exoplanets around their host star with direct
	observations is very challenging and has been enabled in the last decade
	thanks to technological advances in ground-based NIR instruments, adaptive
	optics, and coronagraphy. Direct observations of exoplanets provide a
	powerful complement to indirect detection techniques and enable the
	exploration, thanks to its high sensitivity for wide orbits, of different 
	regions of the parameter space. Direct imaging also allows us to put
	important constraints in planet formation models and planetary systems
	dynamics and, since we obtain the photons from the planets themselves, we
	can proceed with further photometric and spectroscopic characterization 
	\citep{oppenheimer09, derosa16}.
	
	High-contrast direct imaging from the ground presents three main challenges:
	the huge difference in contrast between the host star and its potential 
	companions (typically ranging from $10^{-3}$ to $10^{-10}$), the small
	angular separation between them, and the image degradation caused by the
	Earth's turbulent atmosphere and optical imperfections of the telescope	and
	instruments. These challenges are addressed with a combined effort of 
	coronagraphy, optimized wavefront control, dedicated observing techniques,
	and image post-processing \citep{guyon05, mawet12}.
	
	Among the observing strategies, angular differential imaging \citep[ADI,
	][]{marois06adi} is the most commonly used high-contrast imaging technique,
	in spite of its limitations, and is the focus of this paper. ADI can be
	paired with several post-processing algorithms, such as the least-squares
	based LOCI \citep[locally optimized combination of images,][]{lafren07},
	the maximum likelihood based ANDROMEDA \citep{mugnier09, cantalloube15}, and
	the family of principal component analysis (PCA) based algorithms
	\citep{soummer12,amara12}. Recent algorithms such as LLSG \citep{gomez16}
	aim to decompose the images into low-rank, sparse, and Gaussian-noise terms
	in order to separate the companion signal from the star point-spread
	function (PSF) and speckle field. A common step in any of these approaches
	is the use of a model PSF. Algorithms of different complexities are
	used to build these optimal reference PSFs and enhance the detectability of
	potential planets and disks in the presence of speckle noise. From now on,
	and throughout this paper, we call reference PSF the algorithmically built
	image that we use with differential imaging techniques for subtracting the
	scattered starlight and speckle noise pattern, and enhancing the signal of
	disks and exoplanets. 

	This paper presents a \python\ library for image processing of 
	high-contrast astronomical data: the Vortex Image Processing
	\citep[\vip, ][]{vipascl, vipzenodo} package.  
	\vip\ provides a wide collection of pre- and post-processing
	algorithms and currently supports three high-contrast
	imaging observing techniques: angular, reference-star, and multi-spectral
	differential imaging. The code encompasses not only well-tested and
	efficient implementations of known algorithms but also state-of-the-art new
	approaches to high-contrast imaging tasks. Our library has been designed as
	an instrument-agnostic toolbox featuring a flexible framework where
	functionalities can be plugged in according to the needs of	each particular
	dataset or pipeline. This is accomplished while keeping	\vip\
	easy-to-use and well-documented. Finally, our package is released as
	open-source hoping that it will be useful to the whole high-contrast
	imaging community. 
	
	This paper is organized as follows. Section \ref{sec:secoverview} gives a
	general overview of the design and structure of the \vip. In section
	\ref{sec:secpreproc} we briefly describe the pre-processing and cosmetic 
	functionalities implemented in \vip. Section \ref{sec:secpostproc}
	goes into the details of reference PSF subtraction for ADI data, exploring
	the available post-processing algorithmic approaches in \vip. Section 
	\ref{sec:secfluxpos} describes the photometric and astrometric extraction
	technique implemented in our package. Section \ref{sec:secsensitivity}
	describes the sensitivity limits estimation, and finally section
	\ref{sec:secvipshowcase} showcases \vip\ using on-sky data. 

	\begin{table*}
	\centering
	\caption{\vip\ subpackages}\label{tab:table1}
	\begin{tabular}{ll} 
		\hline \hline
		\noalign{\smallskip}
		Subpackage & General description  \\
		\noalign{\smallskip}
		\hline
		\noalign{\smallskip}
		\texttt{conf}		& Timing, configuration and internal utilities \\
		\texttt{exlib}		& Code borrowed from external sources \\
		\texttt{fits} 		& Fits input/output functionality \\
		\texttt{llsg} 		& Local low-rank + sparse + Gaussian-noise
							  decomposition for ADI data\\
		\texttt{madi} 		& Standard ADI recipe (median PSF reference) \\
		\texttt{negfc} 		& Negative fake companion technique for flux and
							  position estimation \\
		\texttt{nmf}		& Non-negative matrix factorization for ADI data \\
		\texttt{pca}		& PCA-based algorithms for ADI, RDI and mSDI data \\
		\texttt{phot} 		& Signal-to-noise and detection of point-like
							  sources. Contrast curve generation for ADI and
							  RDI data\\
		\texttt{preproc}	& Low-level image operations. Pre-processing and
							  cosmetic procedures \\
		\texttt{stats} 		& Statistics from frames and cubes, correlation and
							  sigma clipping procedures\\
		\texttt{var}		& Filtering, 2d-fitting, shapes extraction and other
							  utilities \\
		\noalign{\smallskip}
		\hline
	\end{tabular}
\end{table*}

\section{Package overview} \label{sec:secoverview}

	The design and development of \vip\ follow modern practices for 
	scientific software development such as code modularity, the active use of
	a version control system (\git) and extensive documentation
	\citep{wilson14}. The code is being developed in \python, and relies on its
	vast ecosystem of scientific open-source libraries/packages including 
	\texttt{numpy} \citep{numpy}, \texttt{scipy} \citep{scipy},
	\texttt{matplotlib} \citep{matplotlib}, \texttt{astropy} \citep{astropy},
	\texttt{scikit-learn} \citep{sklearn}, \texttt{pandas} \citep{pandas} and 
	\texttt{scikit-image} \citep{skimage}. 
	For low-level image processing operations, \vip\
	can optionally use, through its \python\ bindings, \texttt{OpenCV} 
	\citep{opencv}, a fast and robust C/C++ library for computer vision and
	image processing. The latest development version of \vip\ is
	available on GitHub\footnote{\url{http://github.com/vortex-exoplanet/VIP}},
	which is also the platform where users and/or collaborators can report bugs
	and make change requests. Every function and class in \vip\ has its own 
	internal documentation attached describing the aim, arguments (inputs), and
	outputs. The internal documentation is part of the \vip's web 
	documentation\footnote{\url{http://vip.readthedocs.io/en/latest/}}, which
	also provides help in installation and troubleshooting. A tutorial  
	dedicated to ADI, in the form of a Jupyter notebook, is shipped in the main
	repository of \vip. 
	
	The structure of \vip, shown in Table \ref{tab:table1}, is modular and
	allows easy extension and re-utilization of functionalities. The code is
	organized, as any other \python\ library, in subpackages (directories)
	encapsulating modules (\python\ files), which in turn contain the functions
	and classes. It is important noting that \vip\ is not a pipeline per se but
	a library, inspired in well established projects such as \texttt{astropy} 
	or \texttt{scikit-learn}, and does not provide a predefined linear
	workflow. Instead, the user must choose which procedures 
	to use and in which order. The results of \vip's calculations are kept in
	memory or displayed, e.g. in plots or figures, and can be later on saved to
	disk in the form of fits files. In the following paragraphs, we describe
	briefly the most relevant functionalities of each subpackage of \vip.
	
	The subpackage \texttt{fits} includes functions for handling files in the 
	FITS format, through \texttt{Astropy} functionalities. It also includes a
	\python\ class which allows controlling \texttt{SAOImage DS9} windows
	(based on the interface to \texttt{SAOImage DS9} through \texttt{XPA} from
	the \texttt{RO} \python\ package) and displaying \texttt{numpy} arrays.
	Thanks to these functions, \vip\ can be fed from disk with any FITS file
	containing a high-contrast imaging datacube. 

	The subpackage \texttt{phot} includes functionalities such as
	signal-to-noise (S/N) estimation, S/N maps generation, automatic detection
	of point-like sources, fake companion injection in ADI cubes, and
	sensitivity limits computation. For planet S/N calculation, we implement
	the small sample statistics approach proposed by \citet{mawet14ss}, which
	uses a two-sample t-test with statistics computed over resolution elements
	(circular apertures of $\lambda/D$ diameter) instead of considering the
	pixels as statistically independent \citep{mawet14ss,gomez16}. The
	subpackage \texttt{stats} contains functions for computing statistics from
	regions of frames or cubes, sigma filtering of pixels in frames, and for
	computing distance and correlation measures between frames. The subpackage
	\texttt{var} includes image filtering, shapes extraction from images and
	2d-fitting (Gaussian, Moffat) among other functionalities. 
	
	Finally, the subpackage \texttt{preproc} contains low-level image 
	operations and pre-processing functionalities as described
	in the next section, while the subpackages \texttt{llsg}, \texttt{madi},
	\texttt{pca} and \texttt{negfc} contain the post-processing algorithms
	which are described in details in the section \ref{sec:secpostproc} for the
	case of ADI datasets. 
	
\section{Pre-processing} \label{sec:secpreproc}
	
	\vip\ accepts datacubes, or sequence of images stacked in a 3d FITS file,
	that have undergone basic astronomical calibration procedures. These
	procedures, such as flat fielding and dark subtraction, in spite of their
	simplicity, are not included in \vip\ due to the heterogeneity of data
	coming from different observatories. This is a sacrifice that we made in
	order to maintain \vip\ as an instrument-agnostic library. We let the users
	perform these procedures with their own tools or with dedicated instrument
	pipelines. \vip\ requires frames that have been at least successfully flat
	fielded and provides algorithms for any subsequent pre-processing task.
	
	Subpackage \texttt{preproc} contains the functions related to low-level
	image operations, such as resizing, upscaling/pixel binning, shifting,
	rotating and cropping frames. All these functions have a counterpart for
	handling cubes or images sequences. Also, it is possible to temporal
	sub-sample and collapse/combine sequences in a single frame. Combining the
	images can be done via a pixel-wise mean, median or trimmed mean operation
	\citep{brandt13}. 
	
	Pre-processing steps are important when working with high-contrast imaging
	sequences. In ADI sequences, it is critical to have the star at the
	very center of the frames and have them all well aligned. \vip\
	(subpackage \texttt{preproc}), makes it possible to register the frames by
	using 2d-Gaussian or Moffat fits to the data, applying Fourier
	cross-correlation \citep[DFT upsampling method, ][]{guizar08}, computing
	the radon transform \citep{pueyo15} for broadband images, or by fitting the
	position of satellite spots specifically created by ripples on the
	deformable mirror \citep{wertz16}. \vip\ includes procedures for detecting
	bad pixels from images and bad frames from datacubes. Bad pixels are
	replaced with the median of the nearest neighbor pixels inside a square
	window of variable size. Bad frame detection
	is implemented using pixel statistics (i.e. using the pixels in a centered
	annulus at a given radius), frame correlation, or ellipticities of
	point-like sources for detecting and discarding outliers. In general we suggest to discard the bad frames from a sequence before 
	proceeding to the post-processing stage. In certain
	scenarios, sky subtraction might be a desirable step. We implemented in
	\vip\ an algorithm for computing optimal sky background frames, learned
	from the provided sky frames (from the whole sky frames or a subset of its
	pixels by applying a mask), using a PCA-like approach. 

\section{Post-processing} \label{sec:secpostproc}

	\begin{figure}
    \begin{center}
    \includegraphics[width=6cm]{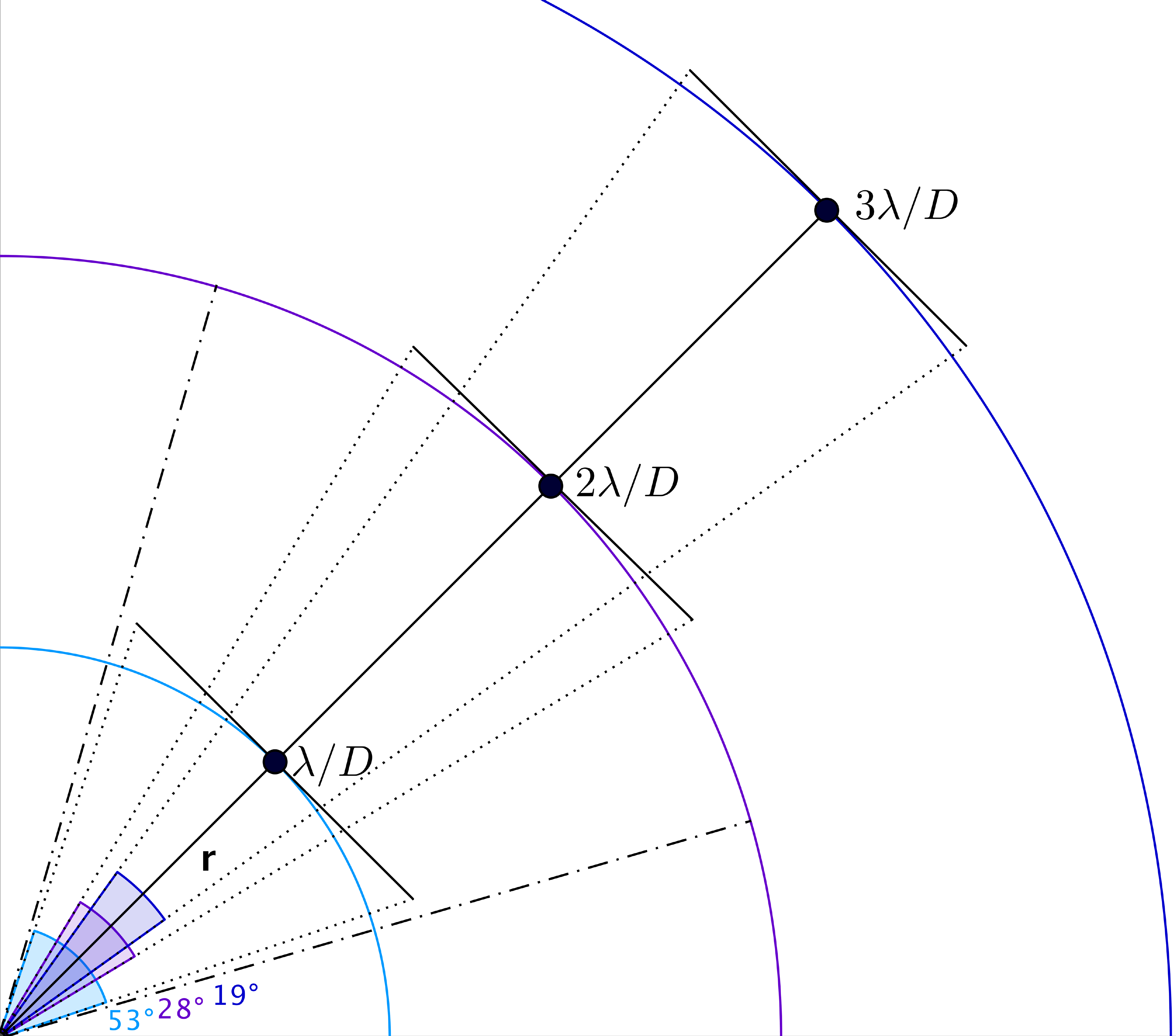} 
    \caption{Illustration of the ADI rotation thresholds at different
	    	 separations in $\lambda/D$. The dot-dashed lines show the
	    	 rejection zone at 2$\lambda/D$ with $\delta=1$ that ensures a
	    	 rotation by at least 1$\times$FWHM ($\lambda/D$) of the PSF.} 
	    	 \label{fig_adi}
    \end{center}
    \end{figure} 	
	
	\subsection{Median reference PSF subtraction}\label{sec:medsubt}
	Subpackage \texttt{madi} contains the implementation of the most basic
	reference PSF subtraction for ADI data \citep{marois06adi}, usually called
	classical ADI in the literature. In this procedure a single reference PSF
	is modeled as the median of the	stack, trying to capture the static and
	quasi-static structures of the sequence. This algorithm can also work in
	annular mode, where an optimized PSF reference can be built for each
	annulus, taking into account a parallactic angle threshold $\omega$ for
	enforcing a given amount of field rotation in the reference frames. The
	threshold $\omega$ is defined as:
	\begin{equation}
	\omega = 2 \arctan{\frac{\delta \cdot \rm FWHM}{2 r}},
	\end{equation}	   
	where FWHM is the Gaussian full width at half maximum in pixels, $\delta$ a
	user-defined threshold parameter, and $r$ the angular separation for which
	the parallactic angle threshold $\omega$ is defined (see Fig. 
	\ref{fig_adi}). The enhanced reference PSF is built for each annulus by
	median combining the $m$ closest in time frames, after discarding neighbor
	frames according to the threshold $\omega$. Median reference PSF
	subtraction has limited performance in the small-angle regime, and it has
	been superseded by more advanced post-processing techniques.		
		
	\subsection{PCA-based algorithms for reference PSF subtraction}
	PCA is an ubiquitous method in statistics and data mining for computing the
	directions of maximal variance from data matrices. It can also be
	understood as a low-rank matrix approximation \citep{absilpa08}. PCA-based
	algorithms for reference PSF subtraction on ADI data can be found in 
	the \vip\ subpackage \texttt{pca}. For ADI-PCA, the reference PSF is
	constructed for each image by projecting the image onto a lower-dimensional
	orthogonal basis extracted from the data itself via PCA. Subtracting from
	each frame its reference PSF produces residual frames where the	signal of
	the moving planets is enhanced. The most basic implementation of ADI-PCA
	uses the whole images by building a matrix $M \in \mathbb{R}^{n\times{p}}$,
	where $n$ is the number of frames and $p$ the number of pixels in a frame.
	The basic structure of the full-frame ADI-PCA algorithm is described in 
	Appendix \ref{appendix:pca_implementation} along with details of \vip's
	implementation. Also, \vip\ implements variations of the full-frame ADI-PCA
	tailored to reduce the computation time and memory consumption when 
	processing big datacubes (tens of GB in memory) without applying temporal
	frame sub-sampling (see Appendix \ref{appendix:pcabigdata} for details).
	Appendix \ref{appendix:pcabigdata} also shows, how temporal sub-sampling can
	degrade the sensitivity of off-axis companions. 

	It is worth noting that in \vip\, a PCA-based algorithm can also be used
	for RDI datacubes, multiple-channel SDI datacubes, and IFS data with
	wavelength and rotational diversities. Data processing for RDI and SDI
	within \vip\ is on-going work and will not be described in more detail in
	this paper.

	\paragraph*{Optimizing $k$ for ADI-PCA.}
	
	\begin{figure}
		\begin{center}
			\includegraphics[width=8cm]{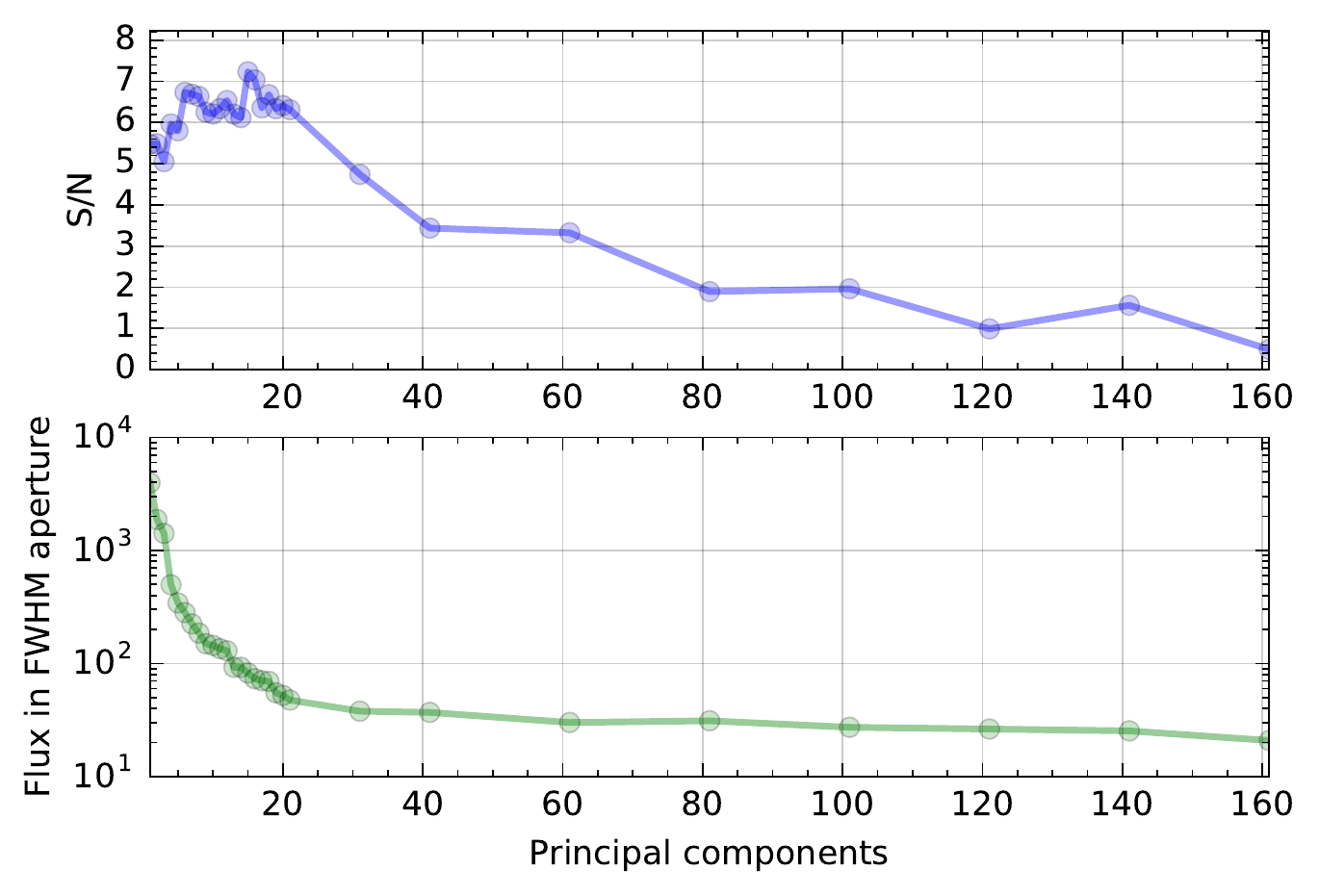}
			\caption{Top: grid optimization of the number of PCs for full-frame 
				ADI-PCA at the location of the HR8799e planet. In this example,
				the mean S/N in a FWHM aperture was maximized with 16 PCs. 
				Bottom: flux of planet HR8799e in a FWHM aperture in the final, post-processed image.}
			\label{fig_pcopt}
		\end{center}
	\end{figure}

	The most critical parameter in every PCA-based algorithms is the number of 
	principal components (PCs) $k$. 
	\vip\ implements an algorithm to find the
	$k$ that maximizes the S/N metric, described in section \ref{sec:secoverview}, for a given location in the image, by running a
	grid search varying the value of $k$ and measuring the S/N for the given
	coordinates. This algorithm can also define an adaptative grid refinement 
	to avoid computing the S/N in regions of the parameter space far from the
	maximum. This algorithm does not deal with the reliability of the
	candidate point-source located at the coordinates of interest. 
	The computational cost remains close to that of a single
	full-frame ADI-PCA run thanks to the fact that we compute the PCA basis
	once with the maximum $k$ we want to explore. Having this basis, we truncate
	it for each $k$ PCs in the grid and proceed to project, subtract and
	produce the final frames where the S/N is computed. An example of such
	optimization procedure is shown in Fig.\ref{fig_pcopt} for the position	
	of planet HR8799e in the dataset described in 
	Sect.\ref{sec:secvipshowcase}. In this case we maximized the mean S/N in a
	FWHM aperture. We can observe that the optimal S/N reaches a plateau near
	the maximum. For true planets, the S/N decreases slowly when increasing the
	number of PCs, as shown in the top panel of Fig.\ref{fig_pcopt}, in
	contrast with a more abrupt S/N decay for noise artifacts or bright
	speckles (which have significant S/N only for a few PCs and quickly fade
	away). The maximum S/N does not correspond to the maximum algorithm
	throughput, and in the illustrated case occurs for a throughput of about
	0.1.    	 	    	
    	
	\paragraph*{Optimizing the library for ADI-PCA.} 
	Full-frame ADI-PCA suffers from companion self-subtraction when the signal
	of interest, especially that of a close-in companion, gets absorbed by the
	PCA-based low-rank approximation that models the reference PSF
	\citep{gomez16}. A natural improvement of this algorithm for minimizing the
	signal loss is the inclusion of a parallactic angle threshold for
	discarding rows from $M$ when learning the reference PSF. This frame
	selection for full-frame ADI-PCA is optional and can be computed for only
	one separation from the star. The idea is to leave in the reference library
	those frames where the planet has rotated by at least an angle $\omega$, as
	described in Sect.\ref{sec:medsubt}. The computational cost increases when
	performing the selection of library frames (for each frame according to its
	index in the ADI sequence) since $n$ singular value decompositions (SVD) 
	need to be computed for	learning the PCs of matrices with less rows than
	$M$. Following the same motivation of refining the PCA library, \vip\
	implements an annular ADI-PCA algorithm, which splits the frames in annular
	regions (optionally in quadrants of annuli) and computes the reference PSF
	for each patch taking into account a parallactic angle rejection threshold
	for each annulus. This ADI-PCA algorithm processes
	$n\times{n_{\rm annuli}}$ (or 4${n}\times{n_{\rm annuli}}$ in case
	quadrants are used) smaller matrices. Details about the annular ADI-PCA
	algorithm implementation can be found in Appendix \ref{appendix:annularpca}.

	\subsection{Non-negative matrix factorization for ADI}
	As previously discussed, the PCA-based low-rank approximation of an ADI
	datacube can be used to model the reference PSF for each one of its frames.
	In the fields of machine learning and computer vision, several approaches
	other than PCA have been proposed to model the low-rank approximation of a
	matrix \citep{kumar16, udell16}. In particular, non-negative matrix
	factorization (NMF) aims to find a $k$-dimensional approximation in terms
	of non-negative factors $W$ and $H$ \citep{leeSeung99}. NMF is
	distinguished from the other methods by its use of non-negativity 
	constraints on the input matrix and on the factors
	obtained. For astronomical images, the positivity is guaranteed since the
	detector pixels store the electronic charge produced by the arriving
	photons. Nevertheless, the sky subtraction operation can lead to negative
	pixels in the background and in this case a solution is to shift all the
	values on the image by the absolute value of the minimum pixel, turning
	negative values into zeros. 	
	
    \begin{figure}
    \begin{center}
    \includegraphics[width=8.5cm]{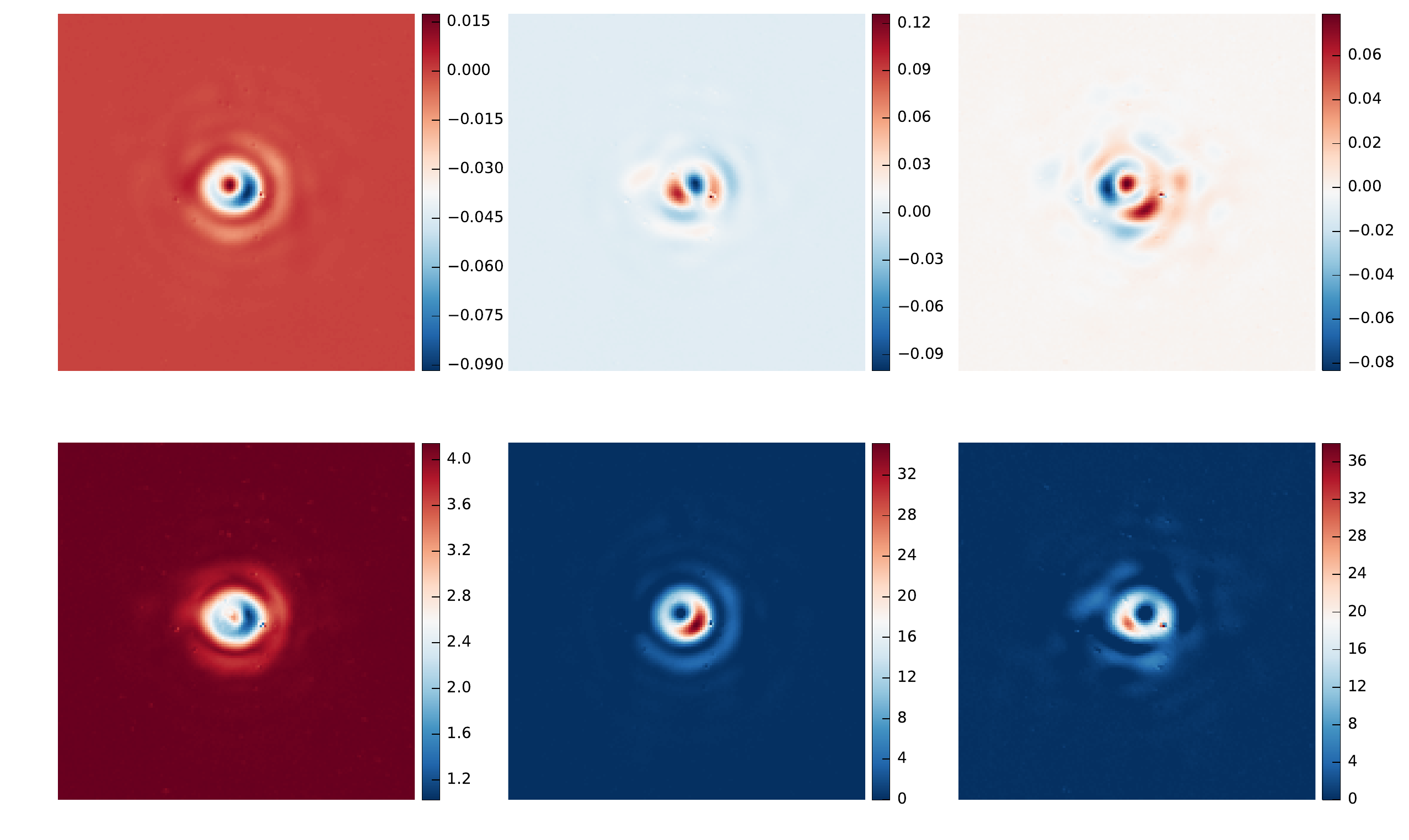}
       \caption{First three principal (top row) and NMF components (bottom row). The components of NMF 
       are strictly positive. This figure is better visualized in color. }
       \label{fig_pcanmf}
    \end{center}
    \end{figure}   	
	
	NMF finds a decomposition of $M$ into two factors of non-negative values,
	by optimizing for the Frobenius norm:
	\begin{equation}
	\argmin_{W,H}{\frac{1}{2}{\left\|M-WH\right\|}_{F}^{2}=\frac{1}{2}\sum_{i,j}{{{(M}_{ij}-{WH}_{ij})}^{2}}},
	\end{equation}
	where $W \in \mathbb{R}^{n\times{k}}$, $H \in \mathbb{R}^{k\times{p}}$, and
	$WH$ is a $k$-rank approximation of $M$. Such a matrix factorization can be
	used to model a low-rank matrix based on the fact that $ \rank(WH) \le
	\mbox{min}(\rank(W),\rank(H))$, where $\rank(X)$ denotes the rank of a
	matrix $X$. Therefore if $k$ is small, $WH$ is low-rank. NMF is a
	non-convex problem, and has no unique minimum. Therefore it is a harder
	computational task than PCA, and no practical algorithm comes with a
	guarantee of optimality \citep{vavasis}. It is worth noting that this
	Frobenius-norm formulation of NMF, as implemented in \texttt{scikit-learn},
	provides final images very similar to the ones from full-frame ADI-PCA. The
	first NMF components along with the first PCs for a same dataset are shown
	in Fig.\ref{fig_pcanmf}. The NMF components are strictly positive and the
	NMF-based low-rank approximation that models the reference PSF for each
	frame is computed as a linear combination of these components. This
	NMF-based algorithm makes a useful complement to PCA-based algorithms for 
	testing the robustness of a detection. 
	    
    \subsection{LLSG for ADI}
    Very recently, a Local Low-rank plus Sparse plus Gaussian-noise
    decomposition \citep[LLSG, ][]{gomez16} was proposed as an approach to
    enhance residual speckle noise suppression and improve the detectability of
    point-like sources in the final image. LLSG builds on recent subspace
    projection techniques and robust subspace models proposed in the computer
    vision literature for the task of background subtraction from natural
    images, such as video sequences \citep{bouwmans14rpca}.	The subpackage
    \texttt{llsg} contains an implementation of the LLSG algorithm for ADI
    datacubes. Compared to the full-frame ADI-PCA algorithm, the LLSG
    decomposition reaches a higher S/N on point-like sources and has overall
    better performance in the receiver operating characteristic (ROC) space.
    The boost in detectability applies mostly to the small inner working angle
    region, where complex speckle noise prevents full-frame ADI-PCA from
    discerning true companions from noise. One important advantage of LLSG is
    that it can process an ADI sequence without increasing too much the
    computational cost compared to the fast full-frame ADI-PCA algorithm. More
    details about LLSG and its performance can be found in \citet{gomez16}.

\section{Flux and position estimation for ADI} \label{sec:secfluxpos}
	\vip\ implements the negative fake companion technique \citep[NEGFC, ][]{marois10,lagrange10} for the determination of the position and flux of
	companions. This implementation is contained in	the subpackage
	\texttt{negfc}. The NEGFC technique consists in injecting in the sequence
	of frames a negative PSF template with the aim of canceling out the
	companion as well as possible in the final post-processed image. The PSF
	template is obtained from off-axis observations of the star. Injecting
	this negative PSF template directly in the images, before they are
	processed, allows to take into account the biases in photometry and
	astrometry induced by the post-processing algorithms. The best cancellation
	of the companion PSF is achieved by minimizing, in an iterative process, a
	well-chosen figure of merit:
	\begin{equation}
	f = \sum_{i=1}^{P}{\abs{I_i}}, 
	\end{equation}
	where $P$ are the pixels contained in a circular aperture (of variable
	radius, by default $4\times{FWHM}$) centered on the companion
	in the final collapsed post-processed image. This NEGFC function of merit
	can be tweaked, by changing the default parameters of the \vip's NEGFC
	procedure. Optionally, one can minimize the standard deviation of the
	pixels, instead of the sum, which according to our test is better in cases
	when the companion is located in a region heavily populated by speckles
	(close to the star), or use the pixels inside a circular aperture from each
	residual frame thus avoiding collapsing the datacube in a single final
	frame. Using the $n\times{P}$ pixels from the residual cube helps getting 
	rid of any bias that the collapsing method, by default a median combination
	of the frames, may introduce.
	
	In \vip\ the estimation of the position and flux (three parameters: radius
	$R$, theta $\theta$ and flux $F$) is obtained by performing	three
	consecutive procedures. A first guess of the flux of a companion is obtained
	by injecting a NEGFC in the calibrated frames while fixing $R$ and $\theta$
	and evaluating the function of merit for a grid of possible fluxes. This
	initial position ($R$ and $\theta$) is determined by visual inspection of
	final post-processed frames or based on prior knowledge. The first guess of
	the companion position/flux can be refined by performing a downhill simplex
	minimization \citep{nelder65}, where the three parameters of the NEGFC are
	allowed to vary simultaneously.
	Although the simplex minimization leads to a significant improvement of the
	position/flux determination, it does not provide error bars on the three
	estimated parameters. More importantly, the function of merit of the NEGFC
	technique is not strictly convex and finding a global minimum is not
	guaranteed. Our approach in \vip\ is to turn our minimization function of
	merit into a likelihood and to use Monte Carlo methods for sampling the
	posterior probability density functions (PDF) for $R$, $\theta$ and $F$.
	This can be achieved via Nested Sampling, calling the \texttt{nestle}
	library \footnote{\url{https://github.com/kbarbary/nestle}}, or Markov
	Chain Monte Carlo (MCMC), through the \texttt{emcee} \citep{emcee} package 
	which implements an affine-invariant ensemble sampler for MCMC. The
	position and flux obtained with the simplex minimization are used for
	initializing the Monte Carlo procedures. From the sampled parameters PDFs,
	we can infer error bars and uncertainties in our estimations, at the cost
	of longer computation time. This NEGFC implementation in \vip\ currently
	works with ADI-PCA-based post-processing algorithms. We also include a
	procedure for estimating the influence of speckles in the astrometric and
	photometric measurements, based on the injection of fake companions at
	various positions in the field of view. More details on the NEGFC
	technique, the definition of the confidence intervals, and details about the
	speckle noise estimation can be found in \citet{wertz16}, where this
	procedure is used to derive robust astrometry for the HR8799 planets.

\section{Sensitivity limits} \label{sec:secsensitivity}

	Sensitivity limits (in terms of planet/star contrast), often referred to as
	contrast curves, are commonly used in the literature for estimating the
	performance of high-contrast direct imaging instruments. They show the
	detectable contrast for point-like sources as a function of the separation
	from the star. \vip\ follows the current practice in building sensitivity 
	curves from \citet{mawet14ss} and thereby applies a student-t correction to
	account for the effect of the small sample statistics. This correction
	imposes a penalty at small separations and therefore the direct comparison
	with contrast curves from previous works may seem more pessimistic close to
	the parent star. The function, contained in the subpackage \texttt{phot},
	requires an ADI dataset, a corresponding instrumental PSF, and the stellar
	aperture photometry $F_{star}$. We suggest removing any real,
	high-significance companions from the data cube (for example using the
	NEGFC approach) before computing a contrast curve.
	The first step is to
	measure the noise as a function of the angular separation $\sigma_R$, on a
	final post-processed frame from the ADI datacube, by computing the standard
	deviation of the fluxes integrated in FWHM apertures. Then we inject fake
	companions to estimate empirically the throughput $T_R$ of the chosen
	algorithm (i.e. the signal attenuation) at each angular separation as 
	$T_R = F_r/F_{in}$, where $F_r$ is the recovered flux of a fake companion
	after the post-processing and $F_{in}$ is the initially injected flux of
	the fake companion. We define the contrast $C_R$ as:
	\begin{equation}
	C_R = \frac{k \cdot \sigma_R}{T_R \cdot F_{star}},
	\end{equation}
	where $k$ is a factor, five in case we want the five sigma contrast curve,
	corrected for the small sample statistics effect. The transmission of the
	instrument, if known, can be optionally included in the contrast 
	calculation.
	
	We note that contrast curves depend on the post-processing algorithm used
	and its tuning, as it will be shown in the panel (a) of Fig.\ref{fig_cc}.
	In signal detection theory, the performance of a detection algorithm is
	quantified using ROC analysis, and several meaningful figures of merit can
	be derived from it \citep{barret06, gomez16}. These figures of merit would
	be better suited than sensitivity curves for post-processing algorithms
	performance comparison. We only provide contrast curve functionality in
	\vip\ as the proper computation and use of ROC curves is subject of
	on-going research.

\section{Applying VIP to on-sky data} \label{sec:secvipshowcase}
	
	We will now proceed to showcase \vip\ on a dataset of HR8799, a young A5V
	star located at 39 pc, hosting a multiple-planet system and a debris disk
	\citep{marois08hr8799}. Its four planets are located at angular separations
	ranging from about $0\farcs4$ to $1\farcs7$, and have masses ranging from 7
	to 10 $M_{J}$ \citep{currie11}. The HR8799 dataset used in the present work
	was obtained at the Large Binocular Telescope \citep[LBT, ][]{hill} on 2013
	October 17, during the first commissioning run of the L’-band annular
	groove phase mask (AGPM) on the LBT Interferometer 
	\citep[LBTI, ][]{hinz, defrere14}.
	A deep ADI sequence on HR8799 was obtained in pupil-stabilized mode on the
	LBTI’s L and M Infrared Camera (LMIRCam), equipped with its AGPM
	coronagraph, using only the left-side aperture. The observing sequence
	lasted for approximately 3.5 hours, providing a field rotation of
	120$\degree$ and resulting in $\sim$17000 frames on target. The seeing was
	fair during the first 30 minutes ($1\farcs2$-$1\farcs4$) and good for the
	remaining of the observations ($0\farcs9$-$1\farcs0$). The adaptive optics
	loop was locked with 200 modes first and with 400 modes after 30 minutes. 
	The off-axis PSF was regularly calibrated during the observations by
	placing the star away from the AGPM center \citep{defrere14}. The raw
	sequence of frames was flat fielded and background subtracted with custom
	routines. The sky background subtraction was performed using the
	median-combination of close in time sky frames. Using a master bad pixel
	mask generated with sky frames taken at the end of the night, the bad
	pixels were subsequently fixed in each frame using the median of adjacent
	pixels \citep{defrere14}.		 
    
	\subsection{Data processing with \texttt{VIP}} 
		
	The calibrated datacube was loaded in memory with \vip\ and re-centered
	using as a point of reference a ghost PSF present in each frame, product of
	a secondary reflection due to the LBTI trichroic 
	\citep{skemer14, defrere14}. The offset between the secondary reflection
	and the central source was measured on the non-saturated PSF observations
	via 2d Gaussian fits. \vip\ includes a function for aligning frames by
	fitting a frame region with a 2d Gaussian profile (using \texttt{Astropy}
	functionality). We used this function for fitting the secondary reflection
	on each frame and placing the star at the center of odd-sized square images
	taking into account the previously calculated offset between the reflection 
	and the main beam.
	
	For the bad frames rejection step, we used the \vip's algorithm based on the
	linear correlation of each frame with respect to a reference from the same
	sequence (30$\times$30 centered sub-frames are used). The reference frame
	was chosen by visual inspection and in agreement with the observing log of
	the adaptive optics system. Ten percent of the frames were finally
	discarded resulting in a datacube with size 15254$\times$391$\times$391
	(after cropping the frames to the region of interest), occupying 9.7 GB of
	disk space (in single float precision).        
    
    The workflow for loading data in memory and pre-processing it with \vip\ is
    as follows:
\begin{lstlisting}
  import vip
    
  # loading the calibrated datacube and 
  # parallactic angles
  cube = vip.fits.open_fits('path_cube')
  pa = vip.fits.open_fits('path_pa')
    
  # aligning the frames
  from vip.calib import cube_recenter_gauss2d_fit\
      as recenter
  cube_rec = recenter(cube, xy=cent_subim_fit, 
      fwhm=fwhm_lbt, subi_size=4, 
      offset=offset_tuple, debug=False)
    
  # identifying bad frames
  from vip.calib import cube_detect_badfr_correlation\ 
      as badfrcorr
  gind, bind = badfrcorr(cube_rec, frame_ref=9628, 
      dist='pearson', percentile=10, plot=False)
      
  # discarding bad frames
  pa_gf = pa[gind]; cube_gf = cube_rec[gind]
    
  # cropping the re-centered frames 
  from vip.calib import cube_crop_frames
  cube_gf_cr = cube_crop_frames(cube_gf, size=391)
\end{lstlisting}

	\begin{figure}
	\begin{center}
		\includegraphics[width=8.5cm]{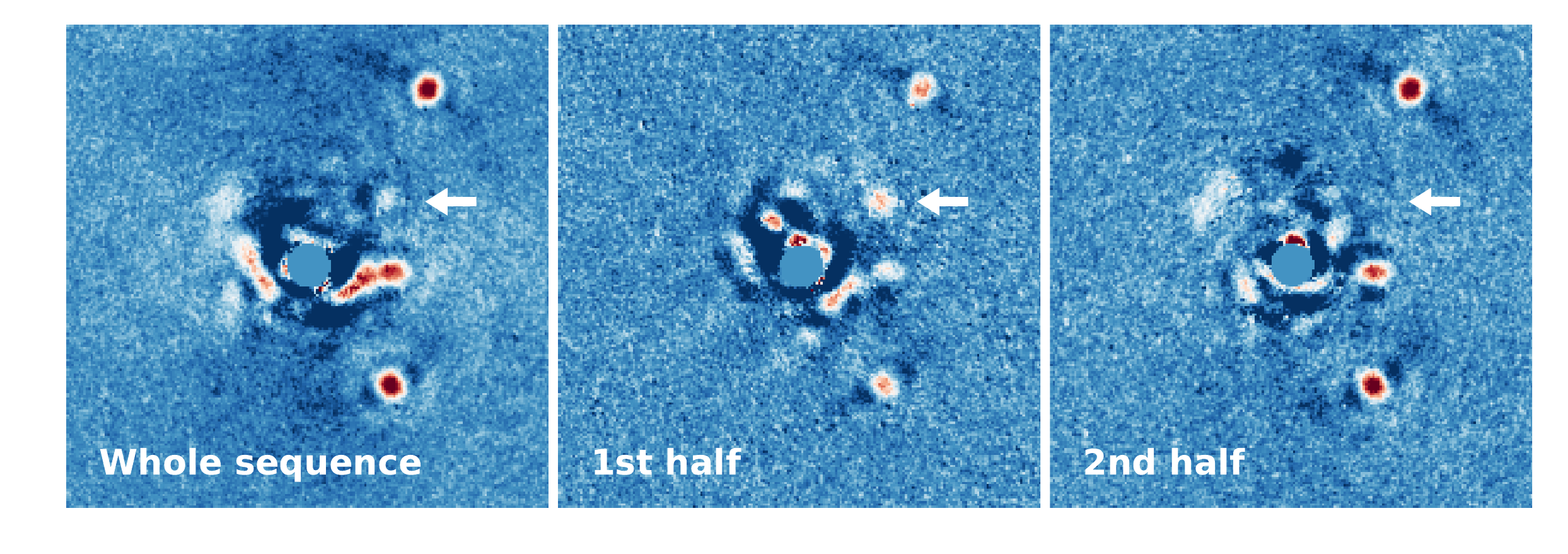}
		\caption{Ghost planet (shown with an arrow) due to the secondary 
			     reflection of LBTI. The left panel shows the full-frame ADI-PCA
		         result using six PCs on the whole ADI sequence. The middle and
			     right panels show the same processing but using only the first
			     and second halves of the sequence.}
		\label{fig_gplan}
	\end{center}
	\end{figure}

	A first exploration of the full-resolution datacube with full-frame ADI-PCA
	showed a feature that resembled an instrumental PSF near the location of
	planet HR8799e, when only a few principal components were used (see left
	panel in Fig.\ref{fig_gplan}).	We concluded, after processing the data in
	two halves, that this blob was a residual artifact of the secondary
	reflection of LBTI, which left a PSF-like footprint due to the slow 
	rotation in the first third of the sequence. The ghost companion appeared
	very bright when using the first half of the sequence and was totally
	absent using the second one, as shown in Fig.\ref{fig_gplan}. Moreover, it
	was located at the same separation as the secondary reflection, whose
	offset was previously measured. For this reason, and because the adaptive
	optics system was locked on 200 modes during the first 5000 frames, while
	it locked on 400 modes for the rest of the sequence, we discarded this
	first batch of frames from the sequence and kept the frames with the
	highest quality. The rotation range of the final sequence is 100$\degree$,
	and the total on-source time amounts to 2.8 hours.	
	
	We processed this datacube with several ADI algorithms, and tuned their
	parameters for obtaining final frames of high quality, where we
	investigated the presence of a potential fifth companion. Other than the
	four known planets around HR8799, we did not find any significant
	detection, worth of further investigation. With the sole purpose of saving
	time while showcasing the \vip\ functionalities, we then decided to
	sub-sample temporally our ADI sequence by mean combining each 20 frames,
	and thereby obtained a datacube of 499 frames. We refrained from binning
	the pixels and worked with the over-sampled LMIRCam images featuring a FWHM
	of nine pixels. The code below illustrates how this steps where done with
	\vip. 
	
	\begin{lstlisting}
	# temporal sub-sampling of frames
	# mean combination of every 20 frames
	from vip.calib import cube_subsample
	cube_ss, pa_ss = cube_subsample(cube_gf[5000:], 
	    n=20, mode='mean', parallactic=pa_gf[5000:])
	
	# cube_ss is a 3d numpy array with shape 
	# [499, 391, 391] and pa_ss a vector [499]
	# with the corresponding paralactic angles
	
	# ADI median subtraction using 2XFWHM annuli, 
	# 4 closest frames and PA threshold of 1 FWHM
	fr_adi = vip.madi.adi(cube_ss, pa_ss, 
	    fwhm=fwhm_lbt, mode='annular', 
	    asize=2, delta_rot=1, nframes=4)
	
	# post-processing using full-frame ADI-PCA
	fr_pca = vip.pca.pca(cube_ss, pa_ss, ncomp=10)
	
	# fr_adi and fr_pca are 2d numpy arrays with 
	# shape [391, 391]
	\end{lstlisting}

	\begin{figure*}
	\begin{center}
		\includegraphics[width=18cm]{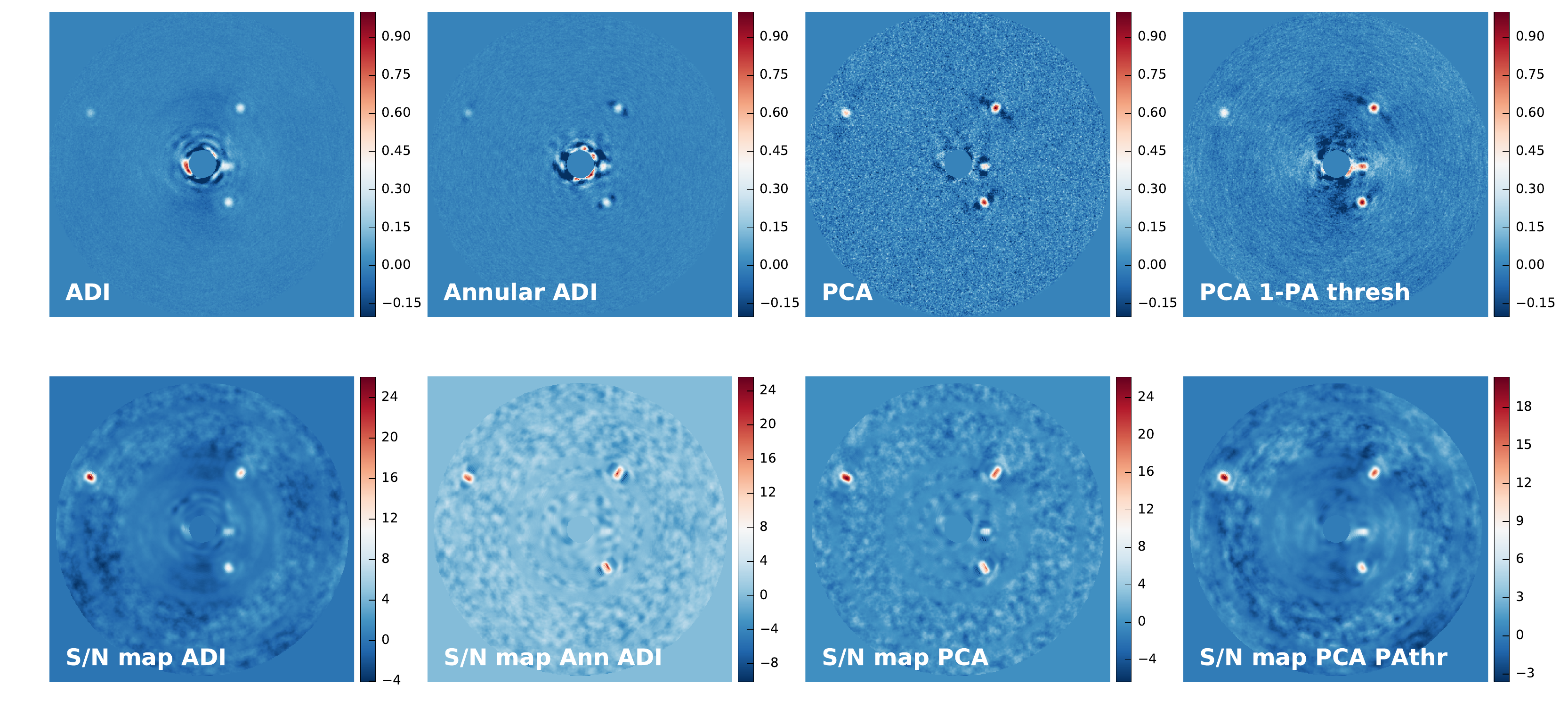}
		\caption{Post-processing final frames (top row) and their corresponding
				 S/N maps (bottom row) for classical ADI, annular ADI,
				 full-frame ADI-PCA and full-frame ADI-PCA with a parallactic
			   	 angle threshold. The final frames have been normalized to their
				 own maximum value. No normalization or scaling was applied to
				 the S/N maps, which feature their full range of values.} 
		\label{fig_pps1}
	\end{center}
	\end{figure*} 	
	
	\begin{figure*}
	\begin{center}
		\includegraphics[width=18cm]{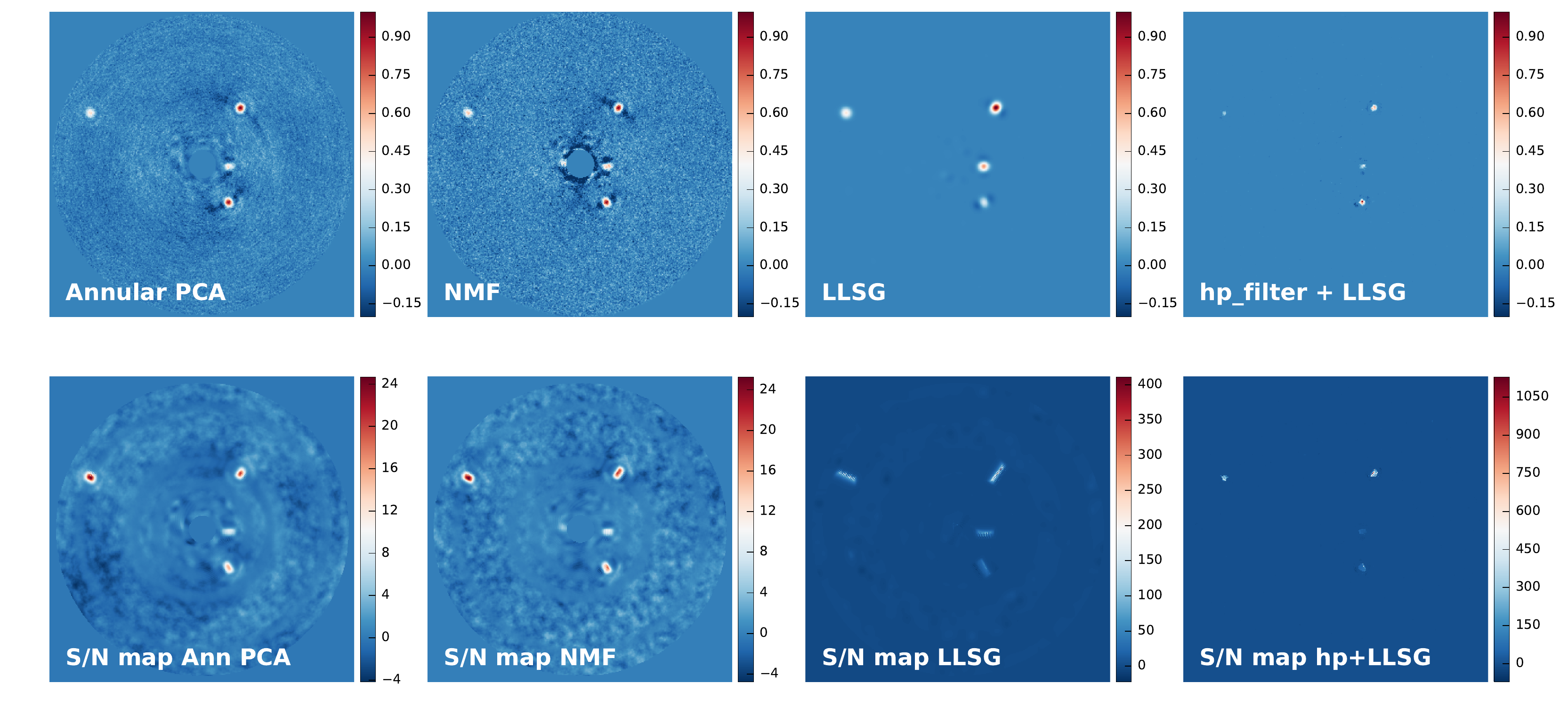}
		\caption{Same as Fig.\ref{fig_pps1} for annular ADI-PCA, full-frame
				 ADI-NMF, LLSG and high-pass filtering coupled with LLSG.
			     We note that a high S/N does not translate in increased 
			     sensitivity to fainter companions.}
		\label{fig_pps2}
	\end{center}
	\end{figure*}

	Figures \ref{fig_pps1} and \ref{fig_pps2} show a non-exhaustive compilation
	of the ADI post-processing options with varying parameters. All the
	algorithms were set to mask the innermost $2\lambda/D$ region.

	We observe how more complex PSF subtraction techniques outperform the 
	classic median subtraction approach for cleaning the innermost part of the
	image ($\sim 2\lambda/D$). We refrain from deriving additional conclusions
	about the comparison of different post-processing techniques as this is 
	beyond the scope of this paper. Furthermore this is an exercise to be
	carried out using a diverse collection of datasets (from different
	instruments) and with appropriate metrics (defined by the whole community),
	such as the area under the ROC curve, in order to provide general and
	robust conclusions. We envision \vip\ as a library that could
	eventually implement all the main high-contrast imaging algorithms and
	become a tool suitable for benchmarking different data processing
	approaches under a unified open framework. 
	
	\subsection{Sensitivity limits and discussion}		

	\begin{figure*}
	\gridline{\fig{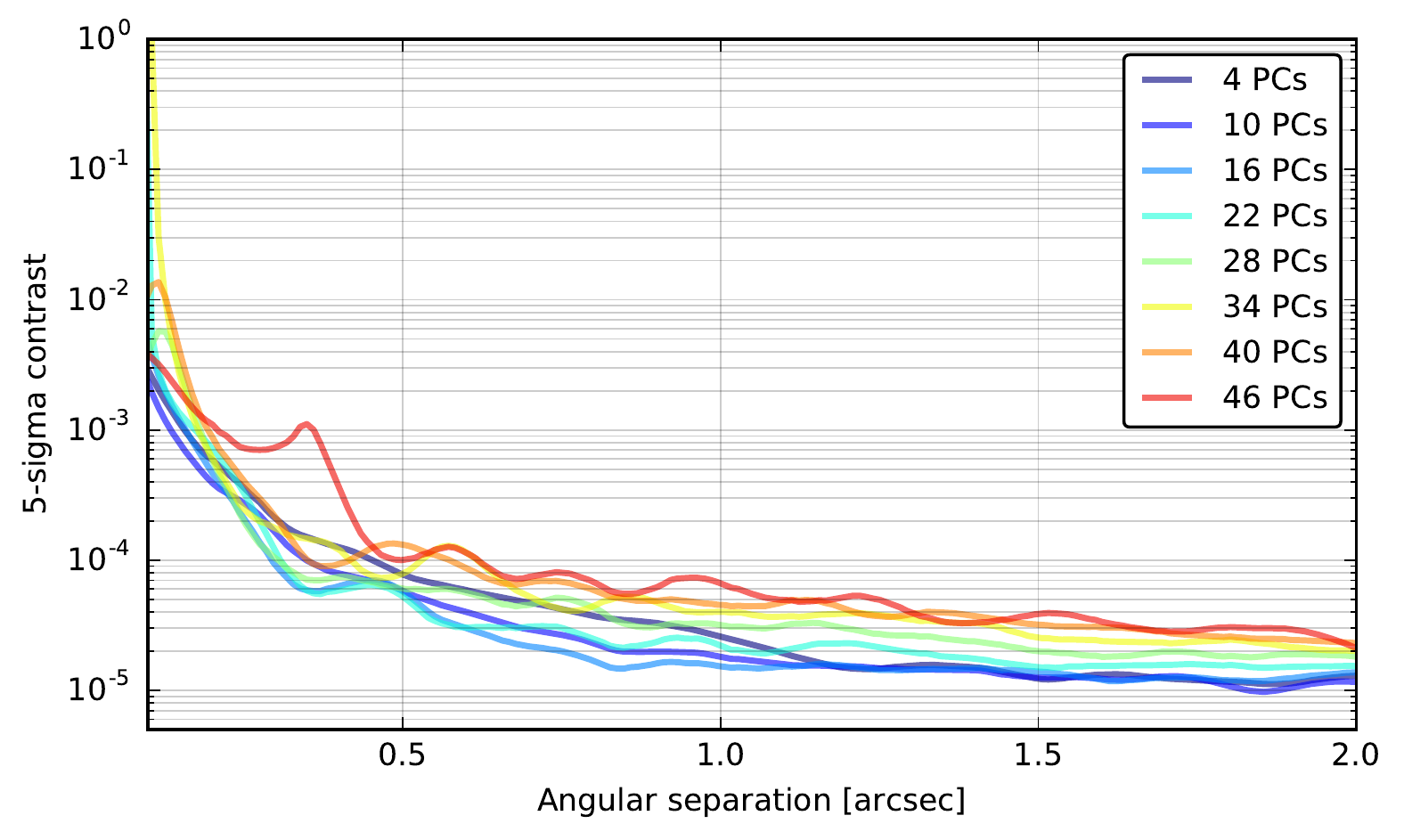}{0.5\textwidth}{(a)}
			  \fig{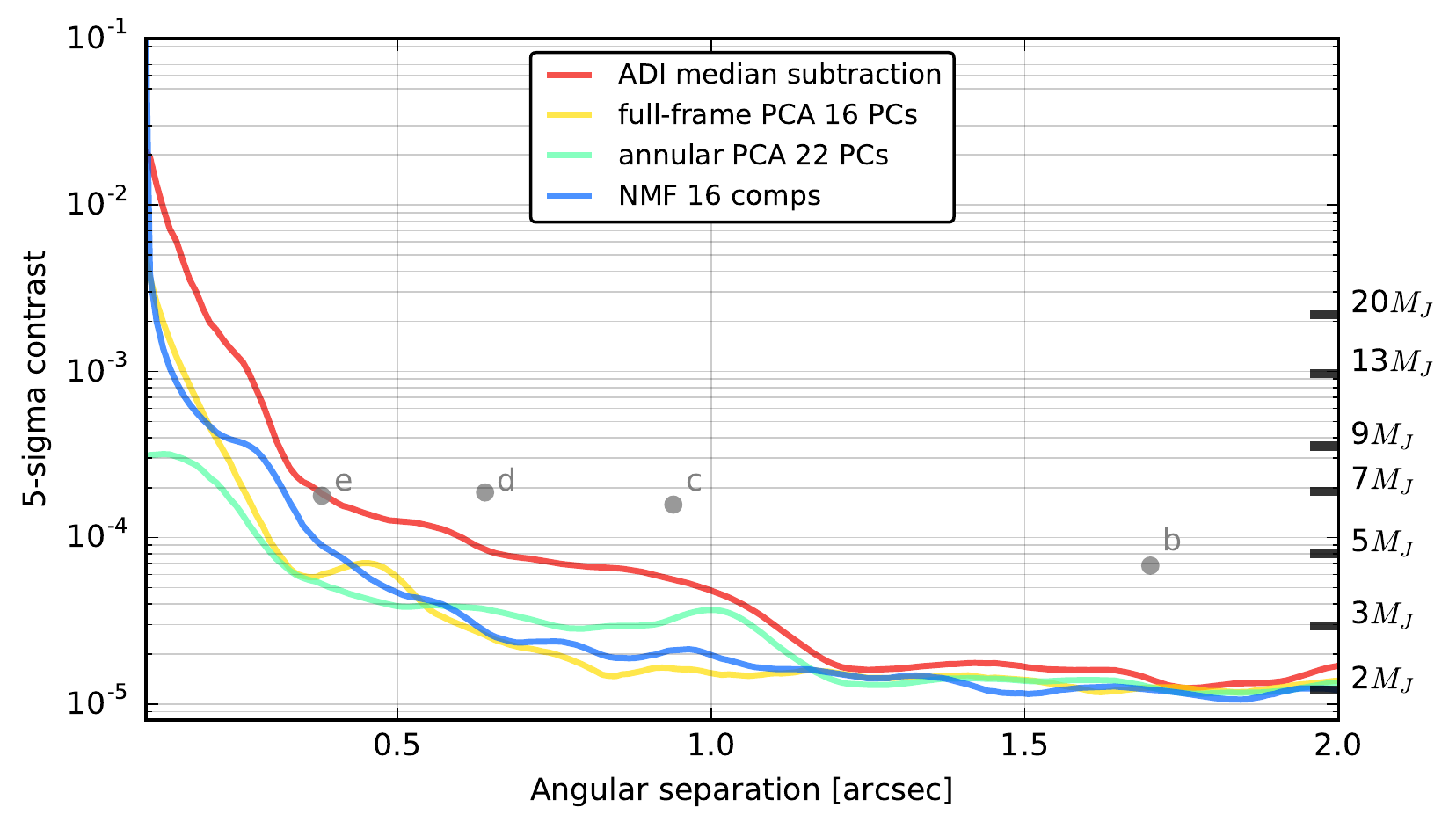}{0.52\textwidth}{(b)}}
	\caption{(a) 5-sigma sensitivity (with the small sample statistics
			 correction) for full-frame ADI-PCA with different numbers of
			 PCs. (b) 5-sigma sensitivities for some of ADI algorithms in
			 \vip. The four known companions were removed before
			 computing these contrast curves. The small sample statistics
		  	 correction has been applied to these sensitivities. A color
		  	 version of this figure is available in the online
		  	 journal.\label{fig_cc}} 	
	\end{figure*}
	
	The code below shows how to compute the S/N for a given resolution element, 
	obtain an S/N map, call the NEGFC MCMC function and compute a contrast
	curve.
	\begin{lstlisting}
	from vip.phot import snr_ss, snrmap, 
	    contrast_curve
	snr_value = snr_ss(fr_pca, source_xy=(54,266), 
	    fwhm=fwhm_lbt)
	
	# S/N map generation
	snr_map = snrmap(fr_pca, fwhm=fwhm_lbt, 
	    plot=True)
	
	# NEGFC mcmc sampling
	from vip.negfc import mcmc_negfc_sampling, 
	    confidence, cube_planet_free
	ini_rad_theta_flux = np.array([r_0, th_0, f_0])
	chain = mcmc_negfc_sampling(cube_ss, pa_ss, 
	    psfn=psf, ncomp=8, plsc=pxscale_lbt, 
	    fwhm=fwhm_lbt, initialState=ini_rad_theta_flux, 
	    nwalkers=100, niteration_min=100, 
	    niteration_limit=400, nproc=10)
	
	# 1-sigma confidence interval calculation
	# from the mcmc chain
	val_max, conf = confidence(chain, cfd=68, 
	    gaussianFit=True, verbose=True)
	final_rad_theta_flux = [(r, theta, f)]
	cube_emp = cube_planet_free(final_rad_theta_flux, 
	    cube_ss, pa_ss, psfn=psf, plsc=pxscale_lbt)
	
	# res_cc is a (pandas) table containing the
	# constrast, the radii where it was evaluated,
	# the algorithmic throughput and other values
	res_cc = contrast_curve(cube_emp, pa_ss, 
	    psf_template=psf, fwhm=fwhm_lbt, 
	    pxscale=pxscale_lbt, starphot=starphot, 
	    sigma=5, nbranch=1, algo=vip.pca.pca, ncomp=8)
	\end{lstlisting}
	
	Using the NEGFC technique, we subtracted the four known companions in our
	datacube and computed the sensitivity curves on this empty datacube. 
	Panel (a) of Fig.\ref{fig_cc} shows the 5-sigma sensitivity for full-frame
	ADI-PCA with varying principal components to exemplify the dependence on
	the algorithm parameters. By using \vip's ADI-PCA algorithm in its annular 
	mode and setting a different number of PCs for each separation, we could 
	obtain the optimal contrast curve, as already shown by \citet{meshkat14}.
	Panel (b) of Fig.\ref{fig_cc} shows the 5-sigma sensitivities for the 
	available ADI algorithms in \vip. These sensitivity limits should be
	representative of the expected performance of the algorithms when applied
	to different data, but the result may vary, therefore preventing us from
	presenting more general conclusions. As expected, in panel (b) of 
	Fig.\ref{fig_cc} we observe how the median reference PSF subtraction
	achieves worse contrast than the rest of the algorithms. Also, we see that
	with annular ADI-PCA, impressive contrast is achieved at small separations
	(below $0\farcs5$) and similar contrast at larger separations if it is
	compared to full-frame ADI-PCA. Annular ADI-PCA presents a smaller
	dependence on the number of principal components (the variance of the
	contrast curves, when varying $k$, is smaller compared to full-frame
	ADI-PCA). For the full-frame ADI-NMF sensitivity, we used 16 components as
	in the case of full-frame ADI-PCA and obtained a fairly similar performance
	at all separations. The contrast metric as defined in \vip\ is not
	well adapted to all	algorithms and/or datasets, therefore we refrain from
	presenting such sensitivity curve for LLSG.
	We remind that these contrast curves were 
	obtained on a temporally sub-sampled datacube. However, because we do not
	include time-smearing when injecting fake companions, we expect the results
	to be representative of the ultimate sensitivity based on the full (non 
	sub-sampled) ADI sequence. The contrast-to-mass conversion for the HR8799
	planets was obtained assuming an age of 40 Myr \citep{bowler16} and using 
	the 2014 version of the PHOENIX BTSETTL models for substellar atmospheric
	models described in \citet{baraffe15}. Based on this, we can discard the
	presence of a fifth planet as bright as HR8799e down to an angular 
	separation of $0\farcs2$. Our detection limits remain in the
	planetary-mass regime down to our inner working angle of $0\farcs1$, and
	reach a background-limited sensitivity of 2$\rm M_J$ beyond about 
	$1\farcs5$.
	
	Finally, it is worth mentioning that the full-frame ADI-PCA sensitivity
	curve presented in this paper (see yellow curve in the panel (b) of 
	Fig.\ref{fig_cc}) is slightly worse than the one shown in \citet{maire15}
	and obtained four days later with the same same instrument but without the
	AGPM coronagraph. In order to make a fair comparison, we re-processed this
	dataset with \vip\ and obtained the same results as \citet{maire15} at
	large angular separations but more pessimistic results closer in (within
	$0\farcs5$). This can be explained by the student-t correction that we
	apply. If we compare the contrast curves produced by \vip\ for both
	datasets, we observe that at small angular separations, within $1 \arcsec$,
	the AGPM coronagraph provides an improvement in contrast up to 1 magnitude.

\section{Summary}

	In this paper we have presented the \vip\ package for data processing of 
	astronomical high-contrast imaging data. It has been successfully tested on
	data coming from a variety of instruments, i.e., Keck/NIRC2, VLT/NACO, 
	VLT/VISIR, VLT/SPHERE and LBT/LMIRCam, thanks to our effort of developing
	\vip\ as an instrument agnostic-library. 
	\vip\ implements functionalities for processing high-contrast imaging data
	at every stage, from pre-processing procedures to contrast curves 
	calculations. Concerning the post-processing capabilities of \vip\, for the
	case of ADI data, it includes several types of low-rank matrix
	approximations for reference PSF subtraction, such as the LLSG
	decomposition, and PCA and NMF-based algorithms. We present, as one of 
	several PCA enhancements, an incremental ADI-PCA algorithm capable of
	processing big, larger-than-memory ADI datasets. In this work we also
	showcased \vip's capabilities for processing ADI data, using a long
	sequence on HR8799 taken with LBTI/LMIRCam in its AGPM coronagraphic mode.
	We used all of \vip's capabilities to investigate the presence of a
	potential fifth companion around HR8799 but we did not find any significant
	additional point-like sources. Further development of \vip\ is planned, in
	order to improve its robustness and efficiency (for supporting big datasets
	in every procedure and multi-processing), and add more state-of-the-art
	algorithms for high-contrast imaging data processing. We propose \vip\ not 
	as a ultimate solution to all high-contrast image processing needs, but as
	an open science exercise hoping that it will attract more users and in turn
	be developed by the high-contrast imaging community as a whole.

\section*{Acknowledgements}

    The authors would like to thank the whole \python\ open-source community 
    and the developers of the powerful open-source stack of scientific libraries. Special thanks to the creators of the Ipython
    Jupyter\footnote{\url{http://jupyter.org}} application. We also thank Elsa
    Huby, Maddalena Reggiani and Rebecca Jensen-Clem for useful discussions and
    early bug reports. Finally, we thank the referee, Tim Brandt, for 
    his constructive questions and valuable comments. The research leading to
 	these results has received
    funding from the European Research Council Under the European Union’s
    Seventh Framework Program (ERC Grant Agreement n. 337569) and from the 
    French Community of Belgium through an ARC grant for Concerted Research 
    Action. V. Christiaens acknowledges financial support provided by 
    Millennium Nucleus grant RC130007 (Chilean Ministry of Economy).
	The LBTI is funded by the National Aeronautics and Space Administration as 
	part of its Exoplanet Exploration Program. The LBT is an international
	collaboration among institutions in the United States, Italy and Germany. 
	LBT Corporation partners are: The University of Arizona on behalf of the
	Arizona university system; Instituto Nazionale di Astrofisica, Italy; LBT 
	Beteiligungsgesellschaft, Germany, representing the Max-Planck Society, the 
	Astrophysical Institute Potsdam, and Heidelberg University; The Ohio State 
	University, and The Research Corporation, on behalf of The University of
	Notre Dame, University of Minnesota and University of Virginia. This 
	research was supported by NASA’s Origins of Solar Systems Program, grant 
	NNX13AJ17G.

\software{numpy \citep{numpy}, scipy \citep{scipy}, matplotlib 	    
	      \citep{matplotlib}, astropy \citep{astropy}, photutils 
	      \citep{photutils}, scikit-learn \citep{sklearn}, pandas \citep{pandas}, scikit-image \citep{skimage},
	      emcee \citep{emcee}, OpenCV \citep{opencv}, SAOImage DS9, nestle}

\appendix
	
\section{Details on the full-frame ADI-PCA algorithm}
	\label{appendix:pca_implementation}
	
	The basic structure of the full-frame ADI-PCA algorithm is the following:
	\begin{enumerate}
		\item the datacube is loaded in memory and $M$ is built by storing on
			  each row a vectorized version of each frame;
		\item optionally $M$ is mean-centered or standardized (mean-centering
			  plus scaling to unit variance);
		\item $k \le \mbox{min}(n,p)$ principal components (PCs) are chosen to
			  form the new basis $B$;
		\item the low-rank approximation of $M$ is obtained as $MB^{T}B$, which
		      models the reference PSF for each frame;
		\item this low-rank approximation is subtracted from $M$ and the result
		      is reshaped into a sequence of frames;
		\item all residual frames are rotated to a common north and are median
		      combined in a final image.
	\end{enumerate}     
	The PCs can be obtained by computing the eigen decomposition (ED) and
	choosing the eigenvectors corresponding to the $k$ largest eigenvalues of
	the covariance matrix $M^{T}M$, or equivalently by computing the SVD of $M$ and extracting the $k$ dominant right
	singular vectors. 
	
	Instead of computing the ED of $M^{T}M$ (which is a large square matrix 
	$p\times{p}$ that must fit in working memory) we can perform the ED of
	$MM^{T}$ for a cheaper PCA computation.	In a similar way, taking the SVD of
	$M^{T}$ is faster and yields the same result as computing the SVD of 
	$M$. Both speed tricks are implemented in \vip. \texttt{Python}, as well as
	other modern programming environments such as \texttt{Mathematica},
	\texttt{R}, \texttt{Julia} and \texttt{Matlab}, relies on \texttt{LAPACK} 
	(Linear Algebra PACKage)\footnote{\url{http://www.netlib.org/lapack/}},
	which is the state-of-the-art implementation of numerical dense linear
	algebra.
	We use the \texttt{Intel MKL} libraries, which provide multi-core optimized 
	high performance \texttt{LAPACK} functionality consistent with the 
	standard. For the SVD, \texttt{LAPACK} implements a ``divide-and-conquer"
	algorithm that splits the task of a big matrix SVD decomposition into some 
	smaller tasks, achieving good performance and high accuracy when working 
	with big matrices (at the expense of a large memory workspace).

\section{Full-frame ADI-PCA for big ADI datasets} \label{appendix:pcabigdata}

	\begin{figure}
		\centering
		\begin{tabular}[b]{@{}p{0.45\textwidth}@{}}
			\centering\includegraphics[width=7.3cm]{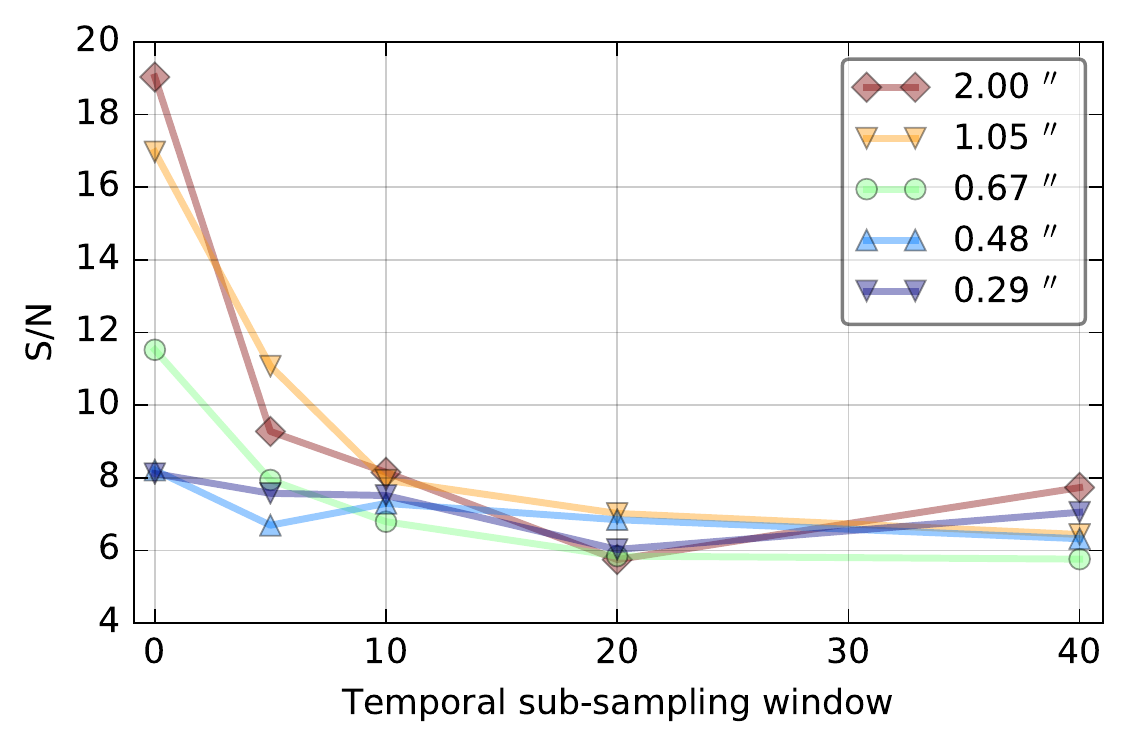} \\
			\centering\small (a) 
		\end{tabular}%
		\quad
		\begin{tabular}[b]{@{}p{0.45\textwidth}@{}}
			\centering\includegraphics[width=7.2cm]{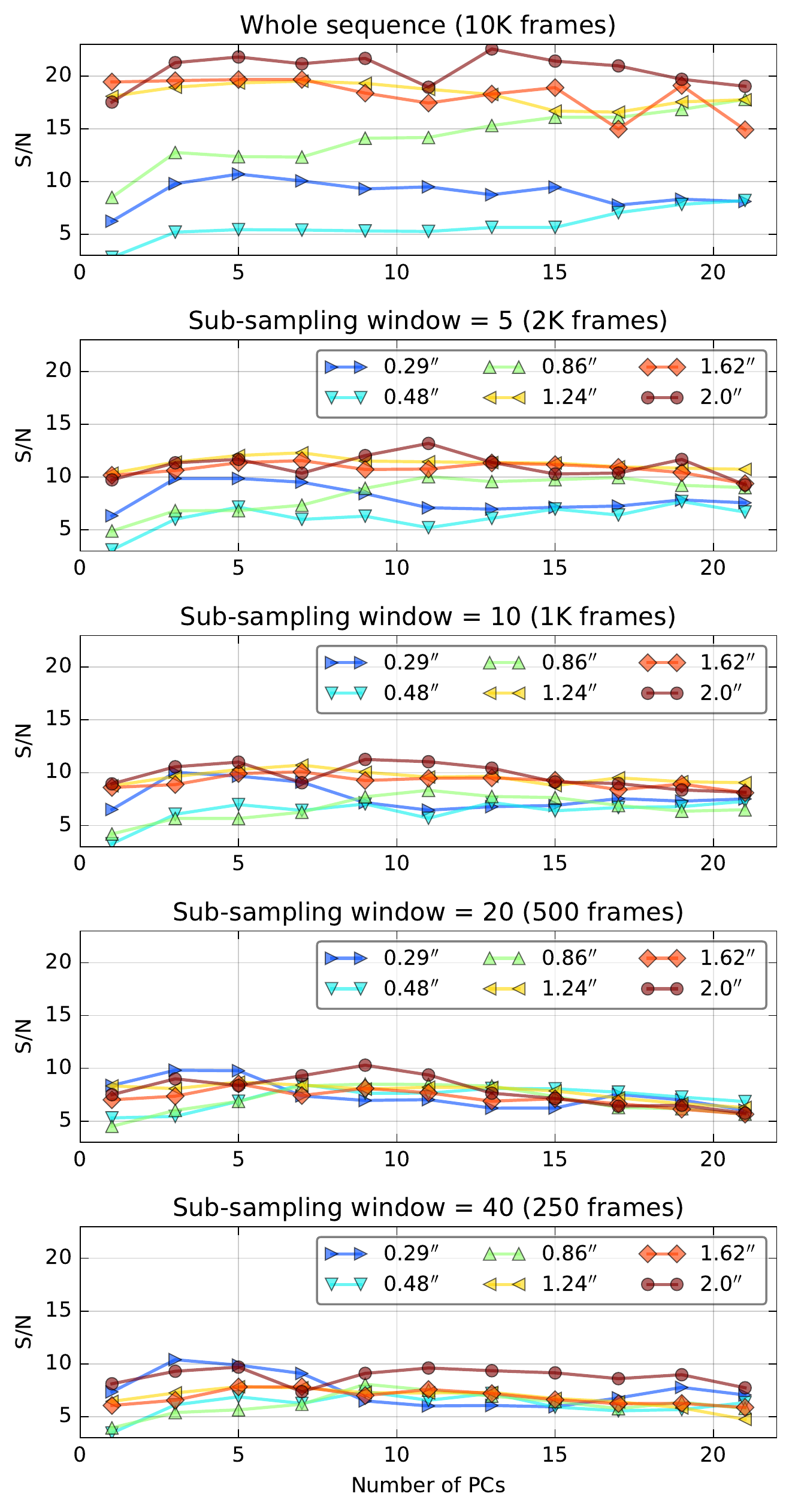}  \\
			\centering\small (b)
		\end{tabular}
		\caption{(a) Fake companions S/N for different angular separations as a
				 function of the temporal sub-sampling applied to the ADI sequence. The horizontal axis shows the amount of frames that were mean combined, with zero meaning that the whole ADI sequence (10k frames) is used. Full-frame ADI-PCA is applied on each datacube, with 21 PCs. 
				 (b) Retrieved S/N on fake companions injected at
				 different angular separations and with a constant flux. The
				 top panel shows the results of varying the number of principal
				 components of the full-frame ADI-PCA algorithm when processing
				 the full resolution ADI sequence. The rest of the panels show
				 the same S/N curves obtained on sub-sampled versions of the
				 sequence using different windows.}
		\label{fig_bigpca}
	\end{figure}

	\begin{figure}
		\centering\includegraphics[width=8.6cm]{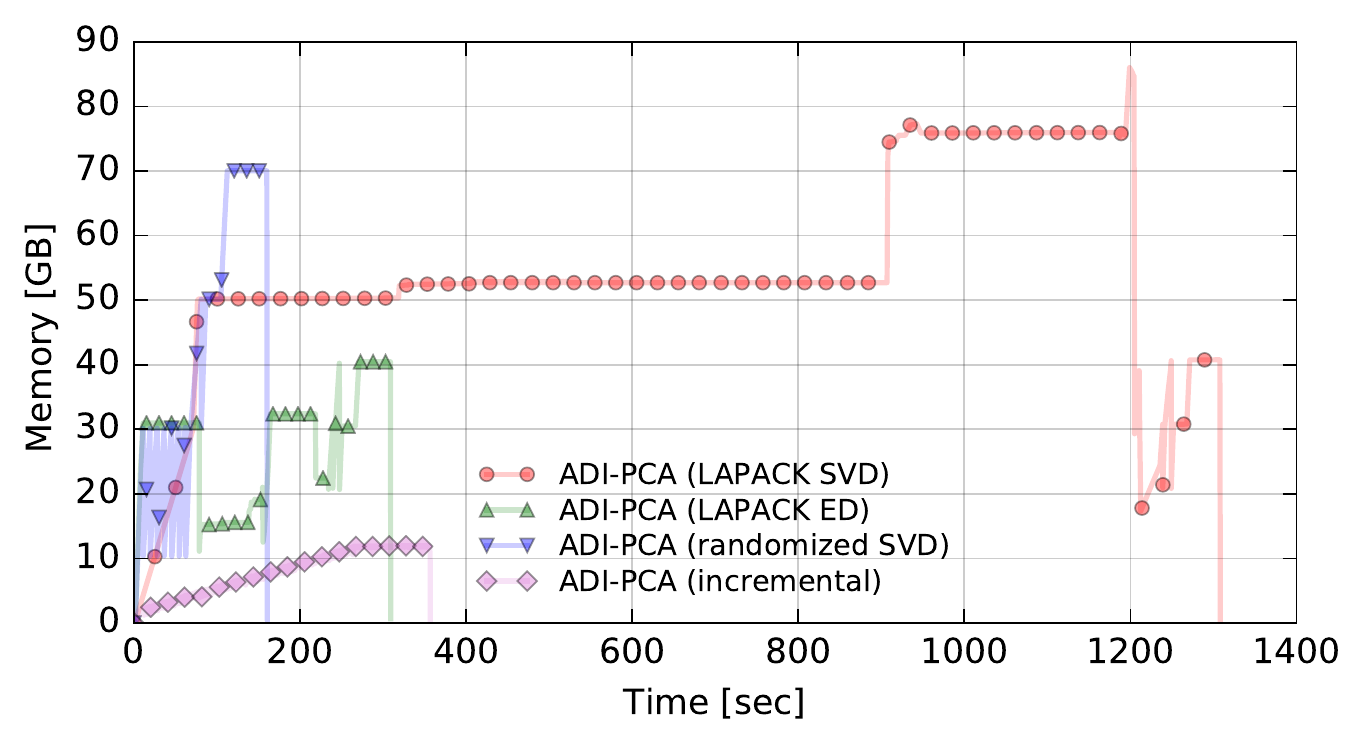} \\
		\caption{Memory usage as a function of the processing time for 	
				 different variations of the full-frame ADI-PCA algorithm on a 
				 large datacube (20 PCs were requested). This is valid for
				 datacubes occupying several GB on disk (in this particular
				 case a 10 GB FITS file was used). It is worth noting that for
				 short or sub-sampled ADI sequences the full-frame ADI-PCA
				 through \texttt{LAPACK} SVD is very efficient and the
				 difference in processing time may become negligible.}
		\label{fig_pca_bench}
	\end{figure}
	
	The size of an ADI dataset may vary from case to case and depends on the 
	observing strategy and the pre-processing steps taken. Typically, a
	datacube contains several tens to several thousands of frames, each one of
	typically 1000$\times$1000 pixels for modern detectors used in 
	high-contrast imaging. In selected instruments (VLT/NACO, LBTI/LMIRCam) 
	that are able to record high-frame rate cubes, a two-hour ADI sequence can
	contain up to $\sim$20000 frames. After cropping down the frames to
	400$\times$400 pixels, we get a datacube in single float values occupying 
	more than 10 GB of disk space. Loading this dataset at once in memory, for
	building $M$, would not be possible on typical personal computer. Even if
	we manage to load the file, the PCA algorithm itself requires more RAM
	memory for SVD/ED calculations, which will eventually cause slowdowns (or 
	system crashes) due to heavy disk swapping. 
	
	The most common approach for dealing with big datasets of this kind is to 
	temporally and/or spatially sub-sample the frames. Reducing the size of the
	dataset effectively reduces the computation time of full-frame ADI-PCA to a
	few seconds but at the cost of smearing out the signal (depending on the
	amount of rotation). Also, depending on the temporal window used for 
	co-adding the frames and on the PSF decorrelation rate, we might end up
	combining sections of the sequence where the PSF has a very different
	structure. It has been stated that there is an optimal window for temporal 
	sub-sampling, which results in increased S/N \citep{meshkat14}. After
	running simulations with fake companion injections at different angular
	separations and measuring the obtained mean S/N in a $\lambda/D$ aperture,
	we came to the conclusion that using the whole sequence of frames (data
	without temporal sub-sampling) is the best choice and delivers the best 
	results in terms of S/N. For this test, we used a datacube of $\sim$10000
	frames, each with 0.5 second of integration time. In panel (a) of 
	Fig.\ref{fig_bigpca} we show the S/N of the recovered companions in
	datacubes sub-sampled using different windows, for an arbitrarily fixed
	number of 21 PCs (even though 21 PCs do not necessarily represent the same
	explained variance for datacubes with different numbers of frames). There
	is an agreement with these results and those in \citet{meshkat14} for
	sub-sampling windows larger than 20, but unfortunately \citet{meshkat14} 
	did not consider smaller sub-sampling windows nor the full ADI sequence.
	More information can be found in panel (b) of Fig.\ref{fig_bigpca}, which
	shows the dependence of the fake companions S/N on the number of PCs for
	different angular separations. Our simulations show very significant gain 
	in S/N when temporal sub-sampling of frames is avoided, especially at large
	separations where smearing effects are the largest.		
	
	\vip\ offers two additional options when it comes to compute the full-frame 
	ADI-PCA through SVD, tailored to reduce the computation time and memory
	consumption when data sub-sampling needs to be avoided (see 
	Fig.\ref{fig_pca_bench}). These variations rely on the machine learning
	library \texttt{Scikit-learn}. The first is ADI-PCA through randomized SVD
	\citep{halko11}, which approximates the SVD of $M$ by using random
	projections to obtain $k$ linearly independent vectors from the range of
	$M$, then uses these vectors to find an orthonormal basis for it and 
	computes the SVD of $M$ projected to this basis. The gain resides in
	computing the deterministic SVD on a matrix smaller than $M$ but with 
	strong bounds on the quality of the approximation \citep{halko11}.
	
	The second variation of the ADI-PCA uses the incremental PCA algorithm proposed by \citet{ross08}, as an extension of the Sequential
	Karhunen-Loeve Transform \citep{levy00}, which operates in on-line
	fashion instead of processing the whole data at once. For the ADI-PCA 
	algorithm through incremental PCA, the FITS file is opened in
	\textit{memmaping} mode, which allows accessing small segments without
	reading the entire file into memory, thus reducing the memory consumption 
	of the procedure. Incremental PCA works by loading the frames in batches of 
	size $b$ and initializes the SVD internal representation of the required
	lower dimensional subspace by computing the SVD of the first batch.
	Then it sequentially updates $n/b$ times the PCs with each new batch until 
	all the data is processed. Once the final PCs are obtained, the same $n/b$ 
	batches are loaded from disk once again and the reconstruction of each 
	batch of frames is obtained and subtracted for obtaining the residuals, 
	which are then de-rotated and median collapsed. A final frame is obtained 
	as the median of the $n/b$ median collapsed frames. A similar approach to 
	incremental PCA, focusing on covariance update, has been proposed by 
	\citet{savransky15}. In Fig.\ref{fig_pca_bench} are compared the 
	memory consumption and total CPU time for all the variants of full-frame 
	ADI-PCA previously discussed. These tests were performed using a 10GB (on 
	disk) sequence and on a workstation with 28 cores and 128 GB of RAM. The 
	results show how incremental PCA is the lightest on memory usage while 
	randomized PCA is the fastest method. 
	With incremental PCA, an appropriate batch size can be used for fitting in
	memory datacubes that otherwise would not, without sacrificing the accuracy 
	of the result. 	

\section{Annular ADI-PCA} \label{appendix:annularpca}

	The annular ADI-PCA comprises the following steps: 
	
	\begin{enumerate}
		\item the datacube is loaded in memory, the annuli are constructed and a parallactic angle
		threshold is computed for each one of them;
		\item for each annulus a matrix $M_{\rm annulus} \in \mathbb{R}^{n\times{p}_{\rm annulus}}$ is 
		built;
		\item optionally $M_{\rm annulus}$ is mean centered or standardized;
		\item for each frame and according to the rotation threshold, a new $M_{\rm opt}$ matrix is 
		formed by removing adjacent rows;
		\item from $M_{\rm opt}$ the $k \le \mbox{min}(n_{\rm opt},p_{\rm annulus})$ principal
		components are chosen to form the new basis optimized for this annulus and this frame;
		\item the low-rank approximation of the annulus patch is computed and subtracted;
		\item the residuals of this patch are stored in a datacube of residuals which is completed
		when all the annuli and frames are processed;
		\item the residual frames are rotated to a common north and median combined in a final image.
	\end{enumerate}  
	
	This algorithm has been implemented with multiprocessing capabilities 
	allowing to distribute the computations on each zone separately. According 
	to our experience, using more than four cores for the SVD computation 
	(through \texttt{LAPACK/MKL}) of small matrices, like the ones we produce 
	in the annular ADI-PCA, does not lead to increased performance due to the 
	overhead in the multi-threading parallelism. When used on a machine with a 
	large number of cores, this algorithm can be set to process each zone in 
	parallel, coupling both parallelisms for higher speed performance. 
	Computing the PCA-based low-rank approximation for smaller patches accounts 
	for different pixel statistics at different parts of the frames. This 
	algorithm can also define automatically the parameter $k$ for each patch by 
	minimizing the standard deviation in the residuals, similar to the
	objective of the original LOCI algorithm \citep{lafren07}, at the expense
	of an increased computation time.

\bibliography{biblio} 
\end{document}